\documentstyle[preprint,revtex]{aps}
\newcommand{\bcn}{\begin{center}}
\newcommand{\beq}{\begin{equation}}
\newcommand{\beqn}{\begin{eqnarray}}
\newcommand{\ecn}{\end{center}}
\newcommand{\eeq}{\end{equation}}
\newcommand{\eeqn}{\end{eqnarray}}

 \def\lsim{\mathrel{\rlap{\lower4pt\hbox{\hskip1pt$\sim$}}
    \raise1pt\hbox{$<$}}}         
\begin{document}
\begin{small}
%
\begin{title}
 $\pi$-$\pi$ scattering in a QCD based model field theory
\end{title}
\author{Craig D. Roberts\footnotemark[2], Reginald T. Cahill\footnotemark[1],\\
Martin E. Sevior\footnotemark[3] and Nicolangelo Iannella\footnotemark[1]
}

\begin{instit}
\footnotemark[2]Physics Division, Argonne National Laboratory, \\
9700 South Cass Avenue, Argonne, Illinois 60439-4843
\end{instit}

\begin{instit}
\footnotemark[1]School of Physical Sciences, Flinders University of South
Australia\\
Bedford Park, SA 5042, Australia
\end{instit}

\begin{instit}
\footnotemark[3]School of Physics, University of Melbourne, \\
Parkville, Vic 3052, Australia
\end{instit}
%
\begin{abstract}
A model field theory, in which the interaction between quarks is mediated by
dressed vector boson exchange, is used to analyse the pionic sector of QCD.  It
is shown that this model, which incorporates dynamical chiral symmetry
breaking,
asymptotic freedom and quark confinement, allows one to calculate $f_\pi$,
$m_\pi$, $r_\pi$ and the partial wave amplitudes in $\pi$-$\pi$ scattering and
obtain good agreement with the experimental data, with the latter being well
described up to energies \mbox{$E\simeq 700$ MeV}.

\footnotetext{~\\
\hspace*{-1.5\parindent}PACS numbers: 12.40.Aa~12.38.Aw~13.75.Lb~14.40.Aq
\hspace*{\fill}Preprint Number: ANL-PHY-7512-TH-93
}

\end{abstract}
%
\section{Introduction}
Chiral Perturbation Theory (ChPT) has been applied extensively and successfully
to a broad range of low energy strong interaction phenomena; both purely
mesonic~\cite{GL83} and with the inclusion of nucleons~\cite{GSS88}.  A
fundamental ingredient of ChPT, and, indeed, what it emphatically establishes,
is the notion that the theory underlying the strong interaction must manifest
dynamically broken chiral symmetry.

Fundamental to obtaining physical information from ChPT, in the sense of
calculating experimental observables, is the fact that at first nonleading
order
[O($E^4$)]; i.e., to fourth order in derivatives of the chiral field, one
obtains a complete effective Lagrangian by including one-pion-loop graphs which
introduce ten counterterms (whose coefficients are infinite).  Higher order
loops do not modify the Lagrangian at this order~\cite{W79} but the ChPT
Lagrangian is only completely defined upon the inclusion of these one-loop
contributions.

This framework can be used to analyse the pionic sector of the strong
interaction: in particular, $\pi$-$\pi$ scattering~\cite{DRV88,BKM91}; and it
is quite successful.  One can choose the coefficients of the counterterms in
such a way as to obtain rather good agreement with the partial wave amplitudes.
The approach of Ref.~\cite{BKM91} augments standard ChPT by the inclusion of
scalar and vector resonances and obtains good agreement with the experimental
S-, P-, D- and F-wave scattering lengths and reasonable agreement with the
phase
shifts up to \mbox{$E\simeq 800$ MeV}.

There is another approach to modelling QCD which we wish to employ to study the
pionic sector of QCD, in general, and $\pi$-$\pi$ scattering in particular.
This approach, as will be seen, actually contains ChPT in the sense that it
generates an effective Lagrangian for the chiral field with the same structure
as that employed in ChPT but with the difference that the coefficients of each
term are finite with values determined by three parameters associated with the
quark-gluon substructure of the model.  Our approach is based on a model
field theory, with elementary quark fields interacting via dressed vector boson
exchange, that has been referred to as the Global Colour-symmetry Model (GCM)
and which was first described in Ref.~\cite{CR85} and extended to the
$\pi$-sector in Ref.~\cite{RCP88}.  The extension to other mesons and to the
baryon is reviewed in Ref.~\cite{RTC92}.

Including quarks explicitly, in the sense of actually carrying out the
operations on the generating functional for the model field theory, \mbox{$\cal
Z$}, that lead to an expression of \mbox{$\cal Z$} in terms of the chiral field
so that the relation between the coefficients in the effective action and those
describing the model quark-gluon dynamics is made explicit, is one property
that
sets this approach apart from ChPT.  Quark confinement is incorporated by
ensuring that the quark propagator obtained in the model has no singularities
that can give rise to absorptive parts associated with free-quark production in
the Feynman diagrams for any given process.  (This is a sufficient but not
necessary condition for quark confinement, as discussed in Ref.~\cite{RWK92}.)

In the GCM the effective action for mesonic degrees of freedom is generated by
a
derivative expansion of the fermion determinant, that arises through an
Hubbard-Stratonovich transformation and subsequent integration over the quark
fields, in a manner analogous to that employed in the Nambu--Jona-Lasinio model
(NJL)~\cite{ER86}.  In contrast to the NJL studies, however, each fermion loop
that arises is finite because the meson fields are not pointlike but, rather,
they are extended objects whose internal structure is given by Bethe-Salpeter
amplitudes obtained via solution of the ladder Bethe-Salpeter
equation.~\cite{CRP87} These Bethe-Salpeter amplitudes play the same role as
the
phenomenological form factors employed in the calculations of intermediate
energy nuclear physics but, it should be stressed, they are calculated
quantities in the GCM.

Another property that sets the GCM apart from ChPT is the fact that meson loops
are also finite.  This is because the meson-meson vertices are not pointlike;
i.e., there are no contact \mbox{$\pi(x)^4$} interactions, but instead each
interaction has a momentum dependence determined by an underlying quark loop
structure.  This quark substructure generates ``form factors'' at the vertices
and therefore the tree level effective action only receives finite corrections
from one-loop diagrams, a fact which is illustrated in Ref.~\cite{HRM92}.  This
is an important difference from ChPT.  The studies of Ref.~\cite{HRM92} also
suggest that, in our model, the the loop corrections are small near threshold.

The GCM is a model which microscopically manifests dynamical chiral symmetry
breaking (DCSB) via the mechanism well know from phenomenological
Schwinger-Dyson equation (SDE) studies~\cite{SDE} and hence it will have in
common with ChPT all of the consequences entailed by this.  Herein we use the
phenomenon of $\pi$-$\pi$ scattering to illustrate this.  Of course, in
specifying a model one has lost the ``model independence'' which is a beauty of
ChPT.  However, one might argue that a stronger connection with QCD has been
obtained since, in addition to DCSB, the GCM incorporates the ultraviolet
renormalisation group features of QCD by construction.  This enables one to
connect observable quantities with the dynamical mass scale of QCD,
$\Lambda_{\rm QCD}$, for example, and also to study their dependence on the
qualitative and quantitative features of the propagators of the elementary
excitations in QCD.

In Sec.~II we briefly summarise the GCM and the procedure that allows one to
proceed to an effective action for the Goldstone modes in the model.  The
action
one obtains is of the same structural form as would be obtained in the NJL
model
but the calculation of physical observables is quite different due to the
internal structure of the mesons.  In this section we also present the
expressions for $f_\pi$, $m_\pi$ and $r_\pi$ in the GCM.  In Sec.~III we study
$\pi$-$\pi$ scattering at tree level.  We calculate the scattering amplitude,
$A(s,t,u)$, and the five lowest order partial wave amplitudes.  In addition,
through exploratory parameter fitting, we demonstrate that the GCM can
reasonably be expected to describe this sector of QCD rather well.  In Sec.~IV
we proceed to use the GCM to calculate the quantities that characterise the
pion
sector.  We present a three parameter version of the model, the parameters
characterising the nature of the quark-quark interaction at small $k^2$, and
show
that a rather good description of the scattering lengths and partial wave
amplitudes is easy to obtain.  We summarise our results and present our
conclusions in Sec.~V.

\section{Effective action for Pions}

The action for the GCM is, in Euclidean metric (with
\mbox{$\{\gamma_\mu,\gamma_\nu\}=2\delta_{\mu\nu}$} and $\delta_{\mu\nu}$ the
Kronecker delta):
\begin{eqnarray}
S[\overline{q},q] & = & \int\,d^4x\,
\overline{q}(x)[\gamma\cdot\partial +M]q(x)
+ \frac{1}{2}\int d^4x\, d^4y\, j_{\mu}^a(x)\,g^2D_{\mu\nu}(x-y)j_{\nu}^a(y)~.
\end{eqnarray}
Here $M$ is the quark mass matrix, \mbox{$j_{\mu}^a(x) =
\overline{q}(x)\frac{\lambda^a}{2}\gamma_\mu q(x)$} and
\begin{equation}
g^2 D_{\mu\nu}(x-y) = \int\,\frac{d^4k}{(2\pi)^4} e^{ik\cdot x}
\frac{1}{k^2}
\left( \left[\delta_{\mu\nu} - \frac{k_\mu k_\nu}{k^2}\right]F_{\rm T}(k^2)
        +\xi \frac{k_\mu k_\nu}{k^2} \right)
\label{eqD}
\end{equation}
is the  dressed gluon propagator in the model.  The specification of
\mbox{$F_{\rm T}(k^2)$} completes the definition of the model.
(It should be noted that the
action for an Abelian theory can be expressed exactly in this form, with
\mbox{$F_{\rm T}(k^2)=1$}, after functional integration over the gauge
field.~\cite{RC86})

The generating functional for the field theory associated with this model
action
can be written:
\begin{equation}
{\cal Z}[\overline{\eta},\eta] =
\int\,{\cal D}\overline{q}{\cal D}q\; \exp\left[-S[\overline{q},q]
        + \int\,d^4x\left[ \overline{\eta}(x)q(x) +
\overline{q}(x)\eta(x)\right]
                \right]
\end{equation}
where \mbox{$\overline{\eta}$} and $\eta$ are Grassmannian sources for the
fermion fields.  Following the bilocal bosonisation procedure described in
detail in Ref.~\cite{RCP88}, a crucial step in which is the application of the
bilocal analogue of the Hubbard-Stratonovich transformation familiar from
statistical mechanics, this generating functional can be rewritten as:
\begin{equation}
{\cal Z}[\overline{\eta},\eta] = \int{\cal D}U\,\exp\left[-S[U]+
{\rm source~terms}\right]
\label{GFU}
\end{equation}
which now involves the local chiral field
\begin{equation}
U(x) = \exp\left( i \frac{T^i \phi^i(x)}{f_\phi} \right)~,
\end{equation}
where \mbox{$\{T^i; i=1,\ldots,N_{\rm f}^2 -1\}$} are the analogue of the Pauli
matrices in \mbox{$SU_{\rm f}(N_{\rm f})$}.

The action in Eq.~(\ref{GFU}) is
\begin{eqnarray}
\label{SOG}
S[U] & = & \Omega[U] +i\Gamma[U]
= -{\rm TrLn}\left[ {\cal G}^{-1}(x,y;[U]) \right] + I[U]
\end{eqnarray}
where
\begin{eqnarray}
\lefteqn{{\cal G}^{-1}(x,y;[U])  =  }\nonumber \\
& & \tilde{G}^{-1}(x-y;\tilde{U})
+ \left\{
P_{\rm R} \left[ U\left(\frac{x+y}{2}\right) - \tilde{U}\right]
+ P_{\rm L} \left[ U^\dagger\left(\frac{x+y}{2}\right) -
\tilde{U}^\dagger\right]
\right\}B(x-y)~,
\label{TrLnarg}
\end{eqnarray}
\begin{equation}
\tilde{G}^{-1}(x-y;\tilde{U}) =
\gamma\cdot\partial\,A(x-y)
+ \left\{P_{\rm R}  \tilde{U}
+ P_{\rm L}  \tilde{U}^\dagger \right\}B(x-y)~,
\end{equation}
and
\begin{eqnarray}
\label{IU}
I[U] & = & \left(\int d^4w\right)
\left(\frac{1}{8N_c} \int d^4z\, {\rm tr}
        \left[
   \left(P_{\rm R} \tilde{U} + P_{\rm L} \tilde{U}^\dagger\right) B(-z)
    T^i\left({\rm tr}\left[T^i \tilde{G}(z;\tilde{U}) \right] +
\right.\right.\right.\nonumber \\
& &\left.\left.\left.
        \gamma_5 {\rm tr}\left[\gamma_5T^i \tilde{G}(z;\tilde{U}) \right]
        \right)\right]
+ \frac{1}{2}\int d^4z\, {\rm tr}\left[\gamma_\mu
\tilde{G}(-z;\tilde{U})\right]
                \partial_\mu\left[ A(z)-\delta^4(z)\right]\right)
\end{eqnarray}
is the term necessary to complete the square in the Hubbard transformation.  In
Eq.~(\ref{IU}),
\begin{eqnarray}
P_{\rm R} = \frac{1}{2}\left(I + \gamma_5\right)~, \;\; & \;\; & \;\;
P_{\rm L} = \frac{1}{2}\left(I - \gamma_5\right)
\end{eqnarray}
are the right and left handed helicity projection operators.

The vacuum or ground state configuration of the chiral field is, as usual,
defined as that configuration \mbox{$\tilde{U}$} for which
\begin{equation}
\left.\frac{\delta}{\delta U} S[U]\right|_{U=\tilde{U}} = 0~.
\label{EL}
\end{equation}
In the present case Eq.~(\ref{EL}) entails that
\mbox{$\tilde{G}^{-1}(z;\tilde{U})$}
%
%
is the solution of the ladder Schwinger-Dyson equation which has the following
form in momentum space:
\begin{equation}
\label{fSDE}
\tilde{G}^{-1}(p;\tilde{U}) = i\gamma\cdot p + M +
\frac{4}{3}\int\frac{d^4k}{(2\pi)^4}
g^2D_{\mu\nu}(p-k) \gamma_\mu \tilde{G}(k;\tilde{U}) \gamma_\nu~.
\end{equation}
If the fermions are massless; i.e., \mbox{$M\equiv0$}, then this (vacuum)
equation admits the same solution for every spacetime independent, unitary
flavour matrix $\tilde{U}$.  This signals that chiral symmetry is dynamically
broken if the ladder SDE admits a nonzero solution for $B(p^2)$; a well known
result seen here from another perspective.

We analyse the action in Eq.~(\ref{GFU}) using a derivative expansion of the
fermion determinant (TrLn).  In this process we expand about the vacuum
configuration $\tilde{U}$ and, with the form of $I[U]$ in Eq.~(\ref{IU}), our
``bare'' pion fields are not pointlike but actually have an internal structure
described by the ladder Bethe-Salpeter equation~\cite{CRP87}: the
Bethe-Salpeter
amplitude and fermion mass function, \mbox{$B(p^2)$}, are identical because of
the identity between the ladder SDE and Bethe-Salpeter equations in vector
exchange theories with dynamically broken chiral symmetry.~\cite{DS79}
We have already made use of this result in writing
Eqs.~(\ref{TrLnarg}-\ref{IU}).

The expansion of the TrLn generates terms whose coefficients are constructed
from quark loops coupling to varying numbers of external pion fields; for
example, the diagram in Fig.~1 is associated with the expression for $f_\pi$.
In this, and the diagrams associated with the other coefficients, the quark
lines represent dressed quark propagators obtained from the solution of
Eq.~(\ref{fSDE}) and the pion Bethe-Salpeter amplitude provides an intrinsic
cutoff on the momentum integration.  As a consequence, each of the coefficients
is finite and the action is completely determined, with no arbitrary finite or
infinite constants being necessary, once the dressed gluon propagator is
specified.  [The only exception to this is the fermion condensate which, in the
more sophisticated versions of the GCM (and QCD itself), diverges with the
upper
bound on the momentum integral in a manner governed by the anomalous dimension
of the fermion propagator so that \mbox{$ m_{\mu^2}^{q}
\langle
\overline{q}q\rangle_{\mu^2}$} is a renormalisation point invariant.  This will
become clear below.]

The real part of the action in  Eq.~(\ref{SOG}) involves even numbers of
pions and, to O($E^4$), it is~\cite{RCP88}:
\begin{eqnarray}
\Omega[U]& = & \int d^4x\, \left\{
\frac{f_{\pi}^2}{4}{\rm tr}\left[\partial_\mu U\partial_\mu U^{\dagger}\right]
+ \frac{\rho_{\mu^2}}{4}{\rm tr}\left[\left(2I_{\rm f} - U - U^{\dagger}\right)
M_{\mu^2}\right]
\right. \\
& & \left. - N_c K_1 {\rm tr}\left[\partial^2 U\partial^2 U^{\dagger}\right]
+ N_c K_2 {\rm tr}\left[
                \left(\partial_\mu U\partial_\mu U^{\dagger}\right)^2\right]
- N_c K_3 \frac{1}{2} {\rm tr}\left[\partial_\mu U\partial_\nu U^{\dagger}
                \partial_\mu U\partial_\nu U^{\dagger}\right]\right\}
\nonumber
\label{RA}
\end{eqnarray}
where \mbox{$M_{\mu^2}= {\rm diag}(m_{\mu^2}^u,m_{\mu^2}^d)$} is the quark mass
matrix (\mbox{$\rho_{\mu^2} M_{\mu^2}$} is renormalisation point invariant in
QCD and in the more sophisticated versions of the GCM) and
\begin{equation}
U(x) = {\rm e}^{ \frac{i}{f_\pi} \,\vec{\tau}\,\cdot\,\vec{\pi}(x)}~.
\end{equation}
At this point we have specialised to the case of $SU_{\rm f}(2)$ [$\vec{\tau}$
are the Pauli matrices] and we have chosen the vacuum configuration
\mbox{$\tilde{U} = I_{\rm f}$} which preserves
\mbox{$SU_{\rm f}^{V}(2)$} symmetry and necessarily follows when \mbox{$M\neq
0$}. The structural form of this part of the action is the same as that of
Ref.~\cite{ER86}.

The imaginary part of the action involves odd numbers of pions and the leading
order term is O($E^5$):
\begin{eqnarray}
i\Gamma[U] & = & \int d^4x\, \frac{N_c}{240\pi^2}\,\epsilon_{\mu\nu\lambda\rho}
\frac{1}{2}{\rm tr}\left[
\left(\tilde{U}^{\dagger}U(x)-U^{\dagger}(x)\tilde{U}\right)
\partial_\mu U(x)\partial_\nu U(x)\partial_\lambda U(x)\partial_\rho U(x)
\right]~.
\label{IA}
\end{eqnarray}
At lowest order in the pion field [O($\pi^5$)] this term yields the Wess-Zumino
term~\cite{WZ71} and when vector mesons are included this imaginary part of the
action contains all of the ``anomalous'' interactions.~\cite{RPC89}

In the GCM, as mentioned above, the coefficients in Eq.~(\ref{RA}) ($f_\pi$,
$\rho$, etc.) are completely determined once the fermion propagator is known
and
this follows as the solution of the fermion SDE, Eq.~(\ref{fSDE}), once the
model gluon propagator, Eq.~(\ref{eqD}), is specified.  Writing the fermion
propagator in the form
\begin{equation}
\tilde{G}(p) = - i \gamma\cdot p \,\sigma_V(p) + \sigma_S(p)
\end{equation}
then:~\cite{RCP88}
\begin{eqnarray}
\lefteqn{f_{\pi}^2  = }\label{Fpi}\\
& & \frac{N_c}{8\pi^2}\int_{0}^\infty\,dx\,B^2\,
\left(  \sigma_{V}^2 -
2 \left[\sigma_S\sigma_S' + x \sigma_{V}\sigma_{V}'\right]
- x \left[\sigma_S\sigma_S''- \left(\sigma_S'\right)^2\right]
- x^2 \left[\sigma_V\sigma_V''- \left(\sigma_V'\right)^2\right]\right)~,
\nonumber\\
\rho_{\mu^2} & = & \frac{N_c}{4\pi^2}\int_{0}^{\mu^2}\,dx\,x\,\sigma_S
\end{eqnarray}
and $K_1$, $K_2$, $K_3$ are given in the appendix.

One can also use this approach to study the electromagnetic pion form factor,
\mbox{$F_\pi(q^2)$}, and this will be described elsewhere.~\cite{CDREPF}  From
the generalised impulse approximation diagram, Fig.~(\ref{RpiGIA}), one obtains
the piece of the charge radius that is regular in the chiral limit,
\mbox{$m_\pi
\rightarrow 0$}, and only weakly dependent on $m_\pi$.  Preliminary studies
suggest that this is the dominant piece at the physical value of
\mbox{$m_\pi$}~\cite{RACDR}, a result supported by the lattice QCD studies of
Ref.~\cite{LC93}.  [In ChPT \mbox{$r_\pi \sim \ln m_{\pi}^2$} as \mbox{$m_\pi
\rightarrow 0$}.  Since the pion is a point
particle in ChPT it derives its form factor from self dressing; the lowest
order
process being interaction with the photon via a pion loop arising from a
\mbox{$\pi\pi \rightarrow \pi\pi$} interaction.  The loop will, of course, have
a branch cut at the mass of the loop particle - in this case $m_\pi$ - and this
is the origin of the logarithmic divergence of $r_\pi$ in ChPT.~\cite{VP74} The
preliminary studies of Ref.~\cite{RACDR} suggest, however, that at the physical
value of $m_\pi$ the charge radius is dominated by the regular contribution
from
Fig.~(\ref{RpiGIA}) which is not present in ChPT.]

\section{$\pi$-$\pi$ Scattering at tree level}
\subsection{Tree level $\pi$-$\pi$ scattering amplitude: \mbox{$A(s,t,u)$}}
In considering $\pi$-$\pi$ scattering we need only consider the real part of
the
action in Eq.~(\ref{RA}).  It will be observed that this differs from that of
Eq.~(5) in Ref.~\cite{DRV88}: in \mbox{$SU_{\rm f}(2)$} the $\alpha_1$ and
$\alpha_2$
terms therein are contained within the $K_{1,2,3}$ terms in our action which
actually has a richer structure.  This is easily seen using the relations:
\begin{equation}
{\rm tr}\left[\left(\partial_\mu U\partial_\mu U^{\dagger}\right)^2\right]
 =  \frac{1}{2}
\left[{\rm tr}\left(\partial_\mu U\partial_\mu U^{\dagger}\right)\right]^2
\end{equation}
and
\begin{eqnarray}
\lefteqn{ {\rm tr}\left[\partial_\mu U\partial_\nu U^{\dagger}
                \partial_\mu U\partial_\nu U^{\dagger}\right]
 =  \frac{1}{2} {\rm tr}\left[\partial_\mu U\partial_\nu U^{\dagger}\right]
 {\rm tr}\left[\partial_\mu U\partial_\nu U^{\dagger}\right]} \\
& - & 2\left[ 2 \partial_\mu S\partial_\mu S \partial_\nu V^i\partial_\nu V^i
 - 2 \partial_\mu S\partial_\nu S \partial_\mu V^i\partial_\nu V^i
 + \partial_\mu V^i\partial_\mu V^i \partial_\nu V^i\partial_\nu V^i
 - \partial_\mu V^i\partial_\nu V^i \partial_\mu V^i\partial_\nu V^i \right]
\nonumber
\end{eqnarray}
where
\begin{eqnarray}
S(x) = \frac{1}{4}{\rm tr}\left[ U(x) + U^\dagger (x)\right] \\
V(x) = \frac{1}{4}{\rm tr}\left[ \tau^i \left(U(x) - U^\dagger
(x)\right)\right]
\end{eqnarray}

As usual, the $\pi$-$\pi$ scattering kernel can be written as follows (with
$s$,
$t$, and $u$ the usual Mandelstam variables):
\begin{equation}
T_{\alpha\beta;\gamma\delta} =
A(s,t,u)\,\delta_{\alpha\beta}\delta_{\gamma\delta}
+ A(t,s,u)\, \delta_{\alpha\gamma}\delta_{\beta\delta}
+ A(u,t,s)\, \delta_{\alpha\delta}\delta_{\beta\gamma}
\end{equation}
which can be decomposed into amplitudes with definite isospin as follows:
\begin{eqnarray}
T^0(s,t,u) & = & 2 \,T_{+-;+-} - 2\,T_{+0;+0} + T_{00;00} =
3 A(s,t,u) + A(t,s,u) + A(u,t,s) \\
T^1(s,t,u) & = & 2\, T_{+0;+0} - T_{++;++}
= A(t,s,u) - A(u,t,s)\\
T^2(s,t,u) & = & 2\, T_{00;00} + 2\,T_{00;+-} + T_{++;++}
= A(t,s,u) + A(u,t,s)~.
\label{Ts}
\end{eqnarray}
The partial wave amplitudes are obtained from these expressions as follows:
\begin{equation}
T_{l}^I(s) = \frac{1}{64\pi}\int_{-1}^{1}dy\, P_l(y) T^I(s,t,u)~.
\end{equation}
(When the pions are on mass shell one has, of course, \mbox{$s+t+u = 4
m_{\pi}^2$} and, in evaluating the partial wave amplitudes it is simplest to
work in the centre of mass frame where \mbox{$t= -(s-4m_{\pi}^2)(1-y)/2$}.)

To calculate $A(s,t,u)$ it is necessary to expand the action of Eq.~(\ref{RA})
to O($\pi(x)^4$):
\begin{eqnarray}
\label{piA}
\Omega[\pi] & = & \int\,d^4x\,\left\{
\frac{1}{2}\left[\partial_\mu \pi^i\partial_\mu \pi^i
+ m_{\pi}^2 \pi^i\pi^i\right]
 - \frac{1}{24}\frac{m_{\pi}^2}{f_{\pi}^2} (\pi^i\pi^i)^2
\right.\\
& &
+ \frac{1}{6f_{\pi}^2}\left[\left(\pi^i\partial_\mu \pi^i\right)^2
                        -\pi^i\pi^i\partial_\mu \pi^j\partial_\mu \pi^j \right]
- \frac{1}{f_{\pi}^4} \left(
 \alpha_1\left(\partial_\mu \pi^i\partial_\mu \pi^i \right)^2
+\alpha_2\, \partial_\mu \pi^i\partial_\nu \pi^i
                \partial_\mu \pi^j\partial_\nu \pi^j \right. \nonumber \\
& &\left.\left.
+ \frac{2}{3} N_c K_1 \left[4
\left(\pi^i\partial^2\pi^i\,\partial_\mu \pi^j\partial_\mu\pi^j
        -\partial_\mu \pi^i\partial^2\pi^i \pi^j\partial_\mu\pi^j
\right)
+ \left(\left(\pi^i\partial^2\pi^i\right)^2
- \pi^i\pi^i \partial^2\pi^j\partial^2\pi^j\right)\right]
\right) \right\}\nonumber
\end{eqnarray}
where
\begin{equation}
f_{\pi}^2 m_{\pi}^2 = \rho_{\mu^2}(m_{\mu^2}^u+m_{\mu^2}^d)
\label{GMOR}
\end{equation}
which is renormalisation point invariant.

In order to facilitate comparisons with Ref.~\cite{DRV88} we have defined
\begin{eqnarray}
\alpha_1 & = & N_c\left(2 K_1 - 2 K_2 -K_3\right)~, \\
\alpha_2 & = & 2 N_c K_3
\end{eqnarray}
in Eq.~(\ref{piA}).

It is now a simple matter to proceed from Eq.~(\ref{piA}) and obtain
\begin{eqnarray}
\label{Astu}
A(s,t,u) & = &
{{{m_{\pi}^2} + 2\,s - t - u}\over {3\,{f_{\pi}^{2}}}}\\
& + & \frac{4 N_c}{3 f_{\pi}^{4}}
\left[ K_1 \left( -12\,{m_{\pi}^4} +
6\,{m_{\pi}^2}\,(s + t +u) + 2\,{s^2} - {t^2}- {u^2}
- 2(\,s\,t +\,s\,u +\,t\,u ) \right)\right. \nonumber \\
& & + K_2 \left.
\left( -2\,{m_{\pi}^2} + s \right) \,\left( -2\,{m_{\pi}^2} + t + u \right)
+ K_3
\left(-2\,{m_{\pi}^4} + {m_{\pi}^2}\,(s +t+u)  -t\,u\right)\right]~. \nonumber
\end{eqnarray}
[It will be remembered that Eq.~(\ref{piA}) is a Euclidean space action.  In
deriving Eq.~(\ref{Astu}) we calculated an algebraic expression for
\mbox{$A(s_{\rm E},t_{\rm E},u_{\rm E})$}
and used this to define an analytic continuation to the physical
\mbox{$(s,t,u)$} domain.]
It is clear that the Weinberg result~\cite{W66} is contained within
Eq.~(\ref{Astu}): one must simply set \mbox{$K_1 = 0 = K_2 = K_3$}; i.e.,
ignore
the terms at O($E^4$).

\subsection{Tree level $\pi$-$\pi$ partial wave amplitudes}
\label{scatfit}
The expressions in Eq.~(\ref{Ts}) can now be used to obtain the scattering
amplitudes at tree level and here we present our calculated expressions for the
first five.  All of these expressions have no imaginary part because we are
working at tree level:
\begin{eqnarray}
T_{0}^0 & = & \nonumber \frac{1}{32 \pi f_{\pi}^2}\left(7m_{\pi}^2 +
\frac{8m_{\pi}^4}{f_{\pi}^2}
            \left[5(\alpha_1+\alpha_2) + \frac{32}{3} N_c K_1\right]
        \right.\label{T00}\\
 & &
 \left.+ 2 x \left[1 + \frac{4 m_{\pi}^2}{f_{\pi}^2}
                        \left[4\alpha_1 + 3\alpha_2 + 4 N_c K_1\right]
             + \frac{x}{3f_{\pi}^2} \left[11\alpha_1 + 7\alpha_2\right] \right]
\right)\\
T_{1}^1 & = &\frac{x}{96\,\pi \,{f_{\pi}^2} }
{\left( 1 + \frac{4 m_{\pi}^2}{f_{\pi}^2}\left[
        (\alpha_2-2\alpha_1+ 4N_c K_1) +
\frac{x}{f_{\pi}^2}(\alpha_2-2\alpha_1)\right]\right)},\\
T_{0}^2 & = & \nonumber \frac{-1}{32 \,\pi\,f_{\pi}^2}\left(2 m_{\pi}^2
        \left[1 - \frac{8m_{\pi}^2}{f_{\pi}^2}
                \left[\alpha_1+\alpha_2 - \frac{8}{3} N_c K_1\right]\right]
        \right.\\
& &
        \left. + x \left[1 -\frac{4m_{\pi}^2}{f_{\pi}^2}
                        \left[ 2\alpha_1 + 3\alpha_2 + 4 N_c K_1 \right]\right]
                - x^2 \frac{4}{3f_{\pi}^2} [\alpha_1+ 2 \alpha_2]\right)\\
T_{2}^0 & = &{{\left( {\it \alpha_1} + 2\,{\it \alpha_2} \right) \,{x^2}}\over
    {240\,\pi \,{f_{\pi}^4}}},\\
T_{2}^2 & = &{{\left( 2\,{\it \alpha_1} + {\it \alpha_2} \right) \,{x^2}}\over
    {480\,\pi\,{f_{\pi}^4} }},
\label{T22}
\end{eqnarray}
where \mbox{$x= s - 4m_{\pi}^2$} and hence threshold corresponds to $x=0$.
The violation of unitarity by these tree level amplitudes will become important
as one moves away from threshold.

Near threshold the partial wave amplitudes have the form:
\begin{equation}
{\rm Re} \left[T_{l}^I (u)\right]=
u^{l}\left( a_{l}^I + u\, b_{l}^I + {\rm O}(u^2)\right),
\end{equation}
where $u=x/(2m_{\pi})^2$, and a first test of the GCM in this sector is the
evaluation of these threshold parameters.  The dimensionless scattering lengths
\begin{equation}
\begin{array}{lllll}
a_{0}^0 = \left.T_{0}^0\right|_{u=0},\; &
a_{1}^1 = \left.\frac{d }{du}T_{1}^1\right|_{u=0},\;  &
a_{0}^2 = \left.T_{0}^2\right|_{u=0},\;
a_{2}^0 = \left.\frac{d^2 }{du^2}T_{2}^0\right|_{u=0},\;  &
a_{2}^2 = \left.\frac{d^2 }{du^2}T_{2}^2\right|_{u=0}
\end{array}
\label{adefs}
\end{equation}
 are not influenced by the imaginary part of the partial wave amplitudes and
hence they provide a useful, simple first test of tree level calculations in a
given model.

The best known of the scattering lengths are $a_{0}^0$ and $a_{0}^2$ and in
Ref.~\cite{S92} it is argued that, experimentally,
\begin{equation}
\label{expas}
\begin{array}{ll}
 a_{0}^0 =  0.20\pm 0.01 ,\;&
a_{0}^2 =-0.037\pm 0.004 .\;
\end{array}
\end{equation}

The others are rather less well known.  In Ref.~\cite{RGDH91}, however, based
on
the experiments of Ref.~\cite{N79}, the following values are presented:
\begin{equation}
\label{datRGDH}
\begin{array}{lll}
 a_{0}^0 =  0.26 \pm 0.005,\;& a_{0}^2 =  -0.028\pm 0.012,\;&
a_{1}^1 =  0.038\pm 0.002,\\
 a_{2}^0 =0.0017 \pm 0.0003,\;&   a_{2}^2 =  0.00013\pm 0.003~. &
\end{array}
\end{equation}
These were the values fitted in Ref.~\cite{BKM91}.

\subsection{Fitting the $\pi$-$\pi$ scattering lengths}
In Ref.~\cite{DRV88}, ChPT was used to fit the scattering lengths: a best fit
being obtained in that parametrisation with:
\begin{eqnarray}
\alpha_1 = -0.0092\;\; & \;& \;\; \alpha_2 = 0.0080
\label{Doalphs}
\end{eqnarray}
which yields:
\begin{equation}
\label{bestD}
\begin{array}{lllll}
 a_{0}^0 =  0.152,\;& a_{0}^2 =  -0.0451,\;&
a_{1}^1 =  0.0363,\, \;  & a_{2}^0 =0.00142,\;&
 a_{2}^2 =  -0.00109~.
\end{array}
\end{equation}

Before proceeding to a calculation of the scattering lengths in the GCM it is
of
interest to determine whether a fit is possible at all and how good it might
be;
i.e., if there are values of the coefficients $K_1$, $K_2$ and $K_3$ in the GCM
which provide a good fit to the scattering lengths.  This is not a well posed
problem since in general, as we have seen, the scattering lengths are not well
known.  For the purpose of illustration we will then simply choose to attempt
to
fit a ``model experimental data set'': $a_{0}^0$ and $a_{0}^2$ from
Eq.~(\ref{expas}); $a_{1}^1$ and $a_{2}^0$ from Eq.~(\ref{datRGDH}); and
$a_{2}^2$ from Eq.~(\ref{bestD}) (based on the plots in Ref.~\cite{DRV88}):
\begin{equation}
\label{ModelExp}
\begin{array}{lllll}
 a_{0}^0 =  0.20,\;& a_{0}^2 =  -0.037,\;&
a_{1}^1 =  0.038,\, \;  & a_{2}^0 =0.0017,\;&
 a_{2}^2 =  -0.0011~.
\end{array}
\end{equation}

We find a best fit (defined as that set of $K$s which minimise the sum of the
squares of the relative difference between the model experimental data and the
calculated values) with
\begin{equation}
\begin{array}{lll}
K_1 = -0.000455,\;  & K_2 = 0.000414,\; & K_3 = 0.00150~,
\end{array}
\label{ExpKs}
\end{equation}
which corresponds to $\alpha_1= -0.00972$ and $\alpha_2=0.00901$,
and these values yield
\begin{equation}
\label{bestas}
\begin{array}{lllll}
 a_{0}^0 =  0.15,\;& a_{0}^2 =   -0.042,\;&
a_{1}^1 =   0.035,\, \;  & a_{2}^0 =0.0017,\;&
 a_{2}^2 =  -0.0011~.
\end{array}
\end{equation}
This fit has a mean deviation in the magnitude of the relative error of 10\%.

It should be remarked at this point that, as far as the at-threshold parameters
are concerned, the only contribution that pion loops will make is to modify the
values of the $K$-parameters calculated from the expressions in the appendix;
i.e., pion loops will only modify the expressions defined implicitly via
Eq.~(\ref{adefs}) by requiring that
\begin{equation}
K^{\rm tree} \rightarrow K^{\rm loops}
\end{equation}
and hence these equations are actually form invariant and may be looked upon as
providing a very general parametrisation of these at-threshold parameters.

If we arbitrarily set $K_1 = 0$ then the expressions in
Eqs.~(\ref{T00}-\ref{T22}) reduce to exactly those of Ref.~\cite{DRV88} (with a
minor correction of a typographical error in \mbox{$T_{0}^0$} therein).  In
this
case the ``best fit values'' of $\alpha_1$ and $\alpha_2$ are:
\begin{eqnarray}
\alpha_1 =   -0.00955\;\; & \; & \;\;\alpha_2 = 0.00897
\end{eqnarray}
which differ little from Eq.~(\ref{Doalphs}) and  correspond to
\begin{equation}
\begin{array}{lll}
K_1 = 0.0,\;  & K_2 =  0.000844,\; & K_3 =   0.00149~.
\end{array}
\label{ExpK10Ks}
\end{equation}
[Obviously, $K_2$ has changed in order to compensate as well as possible for
the
loss of the $K_1$ degree of freedom: In Eq.~(\ref{ExpKs})
\mbox{$K_2 - K_1 = 0.000869$} to be compared with $K_2$ in
Eq.~(\ref{ExpK10Ks}).]  Using Eq.~(\ref{ExpK10Ks}) one obtains:
\begin{equation}
\label{bestasD}
\begin{array}{lllll}
 a_{0}^0 = 0.15 ,\;& a_{0}^2 =-0.045,\;& a_{1}^1 = 0.037,\; & a_{2}^0
=0.0018,\;& a_{2}^2 =-0.0011~.
\end{array}
\end{equation}
This fit has a mean deviation in the magnitude of the relative error of 11\%.
In this illustrative example the extra freedom associated with the coefficient
$K_1$ allows for a marginally improved fit in the GCM.

We have summarised these results in Table.~1.

\section{Model evaluation of the Scattering Lengths}
\label{scatcalc}
\subsection{Point Meson Limit}
We turn now to the direct evaluation of the scattering lengths in the GCM using
the formulae in the appendix that relate the coefficients $K_1$, $K_2$ and
$K_3$
to the quark-gluon substructure in the GCM.  The first and simplest calculation
is the point meson limit. Writing the fermion propagator in the form
\begin{equation}
\tilde{G}(p)= K(p^2) \left(-i\gamma\cdot p A(p^2) + B(p^2)\right)
\end{equation}
with
\begin{equation}
K(p^2) = \frac{1}{p^2 A(p^2)^2 + B(p^2)^2}
\end{equation}
the point meson limit is obtained with
\begin{equation}
A(p^2)=1  \;\;\; {\rm and} \;\;\; B(p^2)= \;{\rm const.}
\end{equation}
This effectively reproduces the results one would
obtain in the NJL~\cite{ER86}.  In this case \mbox{$f_{\pi}$} and
\mbox{$\langle \overline{q} q\rangle$} are both given by divergent momentum
integrals which must be regularised but the $K$s are finite and, in fact,
\begin{equation}
K_1 = K_2 = K_3 = \frac{1}{96 \pi^2}~.
\end{equation}
(This corrects a typographical error in Ref.~\cite{RCP88}.)
In this limit $r_\pi$ is also finite and from Eq.~(\ref{Rpi}) one obtains:
\begin{equation}
r_\pi = \frac{\surd N_c}{2\pi f_\pi}~.
\end{equation}
which agrees with that in the infinite cutoff limit of the NJL
model,~\cite{BHS88} as it should.

In order to present results in this case we simply take the experimental values
of \mbox{$f_{\pi} = 0.093$ GeV} and \mbox{$m_\pi = 0.1385$ GeV} (these values
can
always be arranged in the NJL) and we find that:
\begin{equation}
\label{PML}
\begin{array}{lllll}
r_\pi = 0.58\, {\rm fm}~, & & & & \\
 a_{0}^0 =   0.17 ,\;& a_{0}^2 =-0.048,\;&
a_{1}^1 =  0.036,\;  & a_{2}^0 =0.0020,\;&
 a_{2}^2 = 0.0~.
\end{array}
\end{equation}
One sees that even in the (properly regularised) point meson limit the GCM
provides a reasonably good description of the scattering lengths.

\subsection{Finite Size Pion}
\label{secFSP}
It is simple matter to go beyond the NJL limit of the GCM.  It will be recalled
that in the GCM the fermion propagator is obtained as a solution of the fermion
SDE: Eq.~(\ref{fSDE}); which, in general, is a pair of coupled, nonlinear,
integral equations the solution of which is completely determined once the
model
gluon propagator is specified.  The solution of these equations in
phenomenological models of QCD is an important area of research in its own
right
and has received a good deal of attention~\cite{SDE,SAA91,WKR91,HW91,Rparis}.
The characteristic features of the fermion propagator obtained in all of these
these studies are the same and may be distilled: In the spacelike region, $A$
is
a slowly varying function of $p^2$ with
\mbox{$A(p^2=0)>1$} and \mbox{$A(p^2=\infty)=1$}; $B$ provides most of the
structure in the propagator with \mbox{$B(p^2=0)\approx 1$~GeV} and with the
asymptotic form expected from the Operator Product Expansion~\cite{DP76} and
QCD
Renormalisation Group:
\begin{eqnarray}
b(x)|_{x\sim\infty} & \rightarrow & \frac{4\pi^{2}\lambda}{3}
                          \frac{\kappa}{x(\ln x)^{1-\lambda}}
\label{eq:BOPE}
\end{eqnarray}
with \mbox{$\lambda = 12/(33-2N_{\rm f})$}
[we use \mbox{$N_{\rm f} = 3$} throughout],
$\kappa=(\ln\,\mu^{2})^{-\lambda}<\overline{q} q>_{\mu}$, a dimensionless
renormalisation point invariant, and \mbox{$x= p^2/\Lambda_{\rm QCD}^2$}.

Taking these considerations into account, and following Ref.~\cite{BCP89} in
which fits to various SDE solutions were made, the following model forms of $A$
and $B$, that incorporate all of the physical information and QCD input
available
from the SDE studies, suggest themselves:
\begin{eqnarray}
\label{Amodel}
A(p^2) & = & \frac{2+p^2}{1+p^2} \\
\label{Bmodel}
B(p^2) & = & b_1 \left(1-{\rm tanh}\left[b_2\left(b_3 +
p^2\right)\right]\right)
        + \frac{p^4}{1+p^4} \frac{4\pi^2\lambda}{3}
         \frac{
        \left(\ln[1/\Lambda_{\rm QCD}^2]\right)^{-\lambda}
                \langle\overline{q}q\rangle_{1{\rm GeV}^2}}
     {p^2 \ln\left(b_4 +
{\displaystyle\frac{p^2}{\Lambda_{\rm QCD}^2}}\right)^{1-\lambda}}~.
\end{eqnarray}
(We remark that these fitting forms should only be used at spacelike momenta:
\mbox{$p^2>0$}; i.e, they are useful for interpolating the SDE solutions but
not
for extrapolating them into the complex \mbox{$p^2$} plane.  Studies of the
solution of model fermion SDEs in the complex plane can be found in
Refs.~\cite{BRW92SC}.)

The experimentally determined QCD mass scale, \mbox{$\Lambda_{\rm QCD}$},
introduced into QCD by renormalisation, can be inferred from a number of
experiments and the average value for \mbox{$N_{\rm f} =4$} in the modified
minimal subtraction renormalisation scheme is:~\cite{PDG92}
\begin{equation}
\Lambda_{\rm QCD} = 238 \, \pm \, 30 \, \pm \, 68 \;\; {\rm MeV}~.
\end{equation}
For \mbox{$N_{\rm f} = 3$} this corresponds to
\begin{equation}
\Lambda_{\rm QCD} \simeq 290\; {\rm MeV}
\end{equation}
with a charm quark mass of \mbox{$1.5$~GeV} and this is the value we use
herein.

A reasonable choice for the value of the quark condensate is
\begin{equation}
\rho_{1{\rm GeV}^2} = \langle\overline{u}u\rangle_{1{\rm GeV}^2}=
\langle\overline{d}d\rangle_{1{\rm GeV}^2} =
\langle\overline{q}q\rangle_{1{\rm GeV}^2} = (0.255 {\rm GeV})^3
\end{equation}
which follows from Refs.~\cite{WKR91,SVZ79} and \mbox{$m_\pi$} is then
determined by Eq.~(\ref{GMOR}) once $f_\pi$ is calculated from Eq.~(\ref{Fpi}).
\mbox{$\left[ {\rm We~fix}\;\;
\left(m_{\mu^2=1\,{\rm GeV}^2}^u +
m_{\mu^2=1\,{\rm GeV}^2}^d\right) = 10\,{\rm MeV}.\right]$}

The undetermined parameters in the SDE based model fermion propagator are
$b_1$-$b_4$.  Choosing values for these parameters is completely equivalent to
choosing a model form for the IR (small $k^2$) behaviour of the model gluon
propagator in the GCM.  This is because the UV (large $p^2$) behaviour of $A$
and $B$ is guaranteed by the requirement that the model gluon propagator
manifest asymptotic freedom at large $k^2$; i.e.,
\begin{equation}
\left. F_{\rm T}(k^2)\right|_{k^2 \;\;\mbox{\footnotesize large}}
= \frac{\lambda\pi}{\ln\left(\frac{k^2}{\Lambda_{\rm QCD}^2}\right)}~.
\end{equation}
This is established by the SDE studies of
Refs.~\cite{SDE,SAA91,WKR91,HW91,Rparis}.

In the fit of Ref.~\cite{BCP89}
\begin{equation}
\label{BurdenP}
\begin{array}{lll}
b_1 = 0.7336 \,{\rm GeV} \; & \; b_2= 4.779\, {\rm GeV}^{-2} \;
&\; b_3 = -0.1435 \,{\rm GeV}^2~.
\end{array}
\end{equation}
which were chosen to give the best possible fit to the SDE results of
Ref.~\cite{PCR88} on the domain \mbox{$s=p^2\in [0,1]$ GeV$^2$}.  It is worth
remarking that the parameter $b_4$ in Eq.~(\ref{Bmodel}) is important only in
that it should be greater that one so that the small $p^2$ behaviour is
governed
by the ``tanh'' piece of the model.  As a consequence we set
\begin{equation}
b_4 = 2
\end{equation}
and consider it fixed hereafter since our results are insensitive to increasing
it by two orders of  magnitude.  This is also the reason for the factor
\mbox{$p^4/(1+p^4)$} in Eq.~(\ref{Bmodel}).

The model quark propagator obtained from Eqs.~(\ref{Amodel}) and (\ref{Bmodel})
can be used to calculate $f_\pi$, $m_\pi$, $r_\pi$ and the partial wave
amplitudes.  As a first step we calculated these quantities using the
parameters
of Eq.~(\ref{BurdenP}) which tests the model SDE solution of Ref.~\cite{PCR88}.
This yields
\begin{equation}
\begin{array}{lllll}
f_\pi = 0.074\,{\rm GeV}, \;\; & \;\; m_\pi = 0.17\,{\rm GeV}, \;\;
& r_\pi = 0.72\,{\rm fm}, & &  \\
 a_{0}^0 =   0.37 ,\;& a_{0}^2 =-0.15,\;&
a_{1}^1 =  0.072,\;  & a_{2}^0 =-0.0030,\;&
 a_{2}^2 = -0.0067~.
\end{array}
\end{equation}
The value of $f_\pi$ agrees with that in Ref.~\cite{PCR88} while the other
quantities have not previously been calculated in this model.

These results are not generally satisfactory, a fact that is not surprising
since the parametrisation of the gluon propagator in Ref.~\cite{PCR88} was
simply a ``first guess''; i.e., no systematic attempt was made to improve the
fit to static properties of mesons.  The obvious next step is to determine
whether better agreement with experiment can be obtained simply by varying
$b_1$, $b_2$ and $b_3$.

We allowed $b_1$, $b_2$ and $b_3$ to vary in order to perform a least squares
fit to
\mbox{$f_\pi =0.093$~GeV}, $m_\pi = 0.1385$~GeV,  $r_\pi = 0.66$~fm,
$a_0^0$ - $a_2^2$ from our model experimental data set, Eq.~(\ref{ModelExp}),
and values of $T_0^0(x)$, $T_0^2(x)$ and $T_1^1(x)$ at $x=0.28$~GeV$^2$ taken
from the plots of Ref.~\cite{DRV88}.  This procedure yields
\begin{equation}
\label{aFit}
\begin{array}{lll}
b_1 = 0.5901\, {\rm GeV}, \;\;& \;\; b_2 =  3.0508\, {\rm GeV}^{-2}, \;\;
& \;\; b_3 = -0.2961 {\rm GeV}^2~.
\end{array}
\end{equation}

The mass function obtained with this parameter set is plotted in
Fig.~\ref{PlotB} and, for comparison, we also plot $B(x=p^2)$ obtained with the
parameters of Eq.~(\ref{BurdenP}) in this figure.  It will be observed that
these functions differ only for
\mbox{$|p|/\Lambda_{\rm QCD}
\mathrel{\rlap{\lower4pt\hbox{\hskip1pt$\sim$}}\raise1pt\hbox{$<$}} 4$}; i.e.,
for infrared momenta.  It is important to recall that Eq.~(\ref{BurdenP})
yields
an exact fit to the solution of the fermion SDE obtained when the model gluon
propagator has an integrable IR singularity
\mbox{$\propto \delta^4(k)$}.  This can be interpreted~\cite{WKR91} as a
regularisation of the $1/k^4$ behaviour argued for in
Refs.~\cite{SM79,UBG80,BP88,BP88b}.

The parameters of Eq.~(\ref{aFit}) lead to significantly improved agreement
with
the data:
\begin{equation}
\label{bBfit}
\begin{array}{lllll}
f_\pi = 0.091\,{\rm GeV}, \;\; & \;\; m_\pi = 0.141\,{\rm GeV}, \;\;
& r_\pi = 0.59\,{\rm fm}, & &  \\
 a_{0}^0 =   0.17 ,\;& a_{0}^2 = -0.052,\;&
a_{1}^1 =  0.032,\;  & a_{2}^0 =0.0011,\;&
 a_{2}^2 = -0.00085~.
\end{array}
\end{equation}
(The values of the $K$-coefficients are listed in Table.~I).
It is of interest to note that  \mbox{$r_\pi$} is too small by $\approx$
10\% which, consistent with the lattice analysis of Ref.~\cite{LC93}, allows
for
a small correction from the $\pi$-loop at the physical value of $m_\pi$.

It is clear now that, apart from $m_\pi$, the characteristic quantities of the
pionic sector are a measure of the IR structure of the model gluon propagator;
i.e., the effective interaction between quarks at small $k^2$.  It is worth
emphasising here that this follows because there is a one to one correspondence
between the IR behaviour of the quark propagator and that of the model gluon
propagator.  In fact, in principle it is possible to invert the fermion SDE,
Eq.~(\ref{fSDE}), to obtain the gluon propagator that corresponds to
Eq.~(\ref{bBfit}).  For our purposes herein, however, this is not necessary.
(Although it should be remarked that the form of $A$ and $B$ is consistent with
the presence of an integrable singularity in the model gluon propagator at
\mbox{$k^2 = 0$}.~\cite{CRP87,MN83,WKR91}) The one thing it is necessary to
recall in this connection is that once one has a form for \mbox{$B(p^2)$} then
the form of the pion's Bethe-Salpeter amplitude, $\Gamma$, follows immediately
because of the identity between $B$ and $\Gamma$ in vector exchange theories
with DCSB:~\cite{DS79}
\begin{equation}
\Gamma_\pi(p^2,P^2=0) \propto B(p^2)~.
\end{equation}

To further illustrate the importance of the IR part of the fermion propagator
we
have calculated the characteristic parameters of the $\pi$-sector using
Eq.~(\ref{aFit}) but suppressing the UV tail contribution in
Eq.~(\ref{Bmodel});
i.e., neglecting the
\mbox{$\ln$} term, in which case:
\begin{equation}
\label{noUVcalc}
\begin{array}{lllll}
f_\pi = 0.090\,{\rm GeV}, \;\; &
 r_\pi = 0.58\,{\rm fm}, & & & \\
 a_{0}^0 =   0.18 ,\;& a_{0}^2 = -0.055,\;&
a_{1}^1 =  0.034,\;  & a_{2}^0 =0.0012,\;&
 a_{2}^2 = -0.00094~.
\end{array}
\end{equation}
Clearly, the UV tail only contributes a small amount.  In this connection we
have also studied the sensitivity of our results to variations in $\Lambda_{\rm
QCD}$ in our parametrisation.  Changes of $\pm$ 50\% produce insignificant ($<
1$\%) changes in the characteristic parameters.  This is somewhat artificial,
however, since in reality the coefficients $b_1$-$b_3$ depend on
$\Lambda_{\rm QCD}$ through the solution of the fermion SDE.

We present the partial wave amplitudes of Eqs.~(\ref{T00}-\ref{T22}),
calculated
with the parameter set of Eq.~(\ref{aFit}), in Fig.~\ref{PWA}.  The agreement
with the experimental data, even away from threshold (out to \mbox{$E \simeq
0.7$~GeV}), is quite good.

Clearly, the GCM is able to provide a good qualitative and quantitative
description of the pionic sector of QCD with only three parameters.

\section{Summary and Conclusions}
We have studied the pionic sector of QCD using a model field theory which
incorporates asymptotic freedom, dynamical chiral symmetry breaking (DCSB) and
quark confinement: the Global Colour-symmetry Model (GCM).  Since the GCM
manifests DCSB, the results of chiral perturbation theory (ChPT) are
necessarily
contained within it.  In this connection, the GCM allows for a microscopic
interpretation of the parameters in ChPT; i.e., it relates them to properties
of
the model dressed quark and gluon propagators in the GCM which are the
fundamental ingredients of this model.

In deriving the O($E^4$) action for $\pi$ in the GCM we demonstrated explicitly
that ChPT phenomenology is contained within the GCM since our action contained
that of Ref.~\cite{DRV88} plus an additional term allowed by chiral symmetry.
We showed that, at the simplest level, this extra term, characterised by a
single parameter, enables a marginally improved fit to the $\pi$-$\pi$
scattering lengths.

Calculations in the GCM are tied to the large body of work that has been
completed in the area of phenomenological coupled
Schwinger-Dyson$\leftrightarrow$Bethe-Salpeter equation (SDBSE) studies in QCD
and, in fact, the application of the GCM can be interpreted solely within that
framework.  In the calculation reported herein we used the SDBSE studies to
motivate a form for the quark propagator - a form that incorporated the
physical
input of asymptotic freedom and the operator product expansion; i.e., the large
$p^2$ (UV) behaviour of the propagator was completely determined - so that our
model propagator had only three parameters.  These parameters, which determine
the small $p^2$ behaviour of the propagator, reflect the fact that little is
known about the form of the quark-quark interaction at small $k^2$ (IR).  (The
SDBSE studies indicate that there is a one to one correspondence between the IR
behaviour of the gluon propagator and the IR behaviour of the quark
propagator.)

Using the three parameter form of the GCM we calculated the quantities that
characterise the pionic sector of QCD: $f_\pi$, $m_\pi$, $r_\pi$ and the lowest
five partial wave amplitudes for $\pi$-$\pi$ scattering at tree level.  We
allowed the parameters to vary in order to obtain the best fit possible.  In
Sec.~\ref{secFSP} we demonstrated that the GCM provides a very good description
of this sector of QCD with a good representation of the partial wave amplitudes
well beyond threshold.

It is worth reiterating here that in our study there are no infinite
regularisation parameters, as there are in ChPT, even if we proceed to
incorporate pion loop diagrams.  We have a model, which can also be understood
as a relativistic potential model, in which the parameters are tied to the
behaviour of the quark-quark interaction in the infrared.

The results we have presented herein provide evidence that a ``two-point
model''
of QCD; i.e., a model in which coloured quark currents interact via dressed
gluon exchange (where the dressing models the effect of gluon and ghost
polarisation diagrams on the gluon propagator but where interactions that
depend
on explicit three- and four-gluon vertices are neglected) provides a good
qualitative and quantitative understanding of low energy phenomena in QCD.  It
is worth remarking in this connection that a better experimental determination
of the scattering lengths would allow for more stringent constraints to be
placed upon the nature of the effective interaction between quarks at small
$k^2$.

\thanks
Some of the calculations described herein were carried out using a grant
of computer time and the resources of the National Energy Research
Supercomputer
Center. The work of CDR was supported by the US Department of Energy, Nuclear
Physics Division, under contract number W-31-109-ENG-38, with additional
funding from the US National Science Foundation, Division of International
Programs, under grant number INT-92~15223.  The work of RTC was supported in
part by the Australian Department of Industry, Technology and Commerce.


\unletteredappendix{Coefficients in  $\Omega[U]$}
In this appendix we present the expression for $r_\pi$ and expressions for the
coefficients $K_1$, $K_2$ and $K_3$ in Eq.~(\ref{RA}).  The manner in which
they
are calculated is described in Ref.~\cite{RCP88} and the expressions presented
here correct some typographical errors therein.  Further, they extend those
expressions by incorporating the dependence on the derivatives of
\mbox{$A(p^2)$}.

We have
\begin{eqnarray}
K_1 & = & - b + c - 2 h + f + g \\
K_2 & =&  a_{K_2} - d + 2 f + g + i - j - 2 l + 2 m + 2 n_{K_2} \\
K_3 & = & a_{K_3} + e - 2 g - 2 i + 4 k + 2 l - 2 m + 2 n_{K_3}
\end{eqnarray}
where
\begin{eqnarray}
a_{K_2} & = & \int\frac{d^4q}{(2\pi)^4}\;
K^4 B^4 \left( A^4 + 2 q^2 A^2 (A A' + q^2 [A']^2)
                + \frac{2}{3} q^4 (A A' + q^2 [A']^2)^2 \right)~,\\
a_{K_3} & = & \int\frac{d^4q}{(2\pi)^4}\;
K^4 B^4 \left( A^4 + 2 q^2 A^2 (A A' + q^2 [A']^2)
                - \frac{2}{3} q^4 (A A' + q^2 [A']^2)^2 \right)~,\\
b & = & \frac{1}{2} \int\frac{d^4q}{(2\pi)^4}\;
B^2 \left( A K A' + A^2 K K' + \right.\nonumber \\
 &&       \frac{q^2}{2} ( K^2 [A']^2 + 2 A K A' K' - A^2 [K']^2 + 2 A K^2 A''
                        + A^2 K K'' ) + \nonumber \\
 &&       q^4 (K [A']^2 K' - A A' [K']^2 + \frac{3}{2} K^2 A' A'' + A K A' K''
                        + \frac{1}{6} A K^2 A''' ) + \nonumber \\
&&
\left. q^6 ( - [A']^2 [K']^2 + K [A']^2 K'' + \frac{1}{3} K^2 A' A''' )
\right)~,\\
c & = & \frac{1}{16} \int\frac{d^4q}{(2\pi)^4}\;
        B^2 \left( K'' + q^2 K''' +  \frac{1}{6} q^4 K'''' \right)~, \\
d & = &  \frac{1}{4} \int\frac{d^4q}{(2\pi)^4}\;\left(
       K^2 ( - (B B')^2 + q^2 B B' ( B B'' - [B']^2 ) )\right. \nonumber\\
& &    \left.    +\frac{1}{3}\, q^4\, ( 4 (K K'' - K'^2) (B B')^2
                   + K^2 ( 2 B^2 B' B'''- (B B'')^2 - [B']^4 ) )\right) \\
e & = & \frac{1}{2} \int\frac{d^4q}{(2\pi)^4}\; q^2 \left(
          K ( 2 K' (B B')^2 + K B^2 B' B'' + K B [B']^3 )\right.\nonumber \\
& &        +\frac{1}{6}\, q^2\,\left.( 4 (K K'' - K'^2) (B B')^2
                   + K^2 ( 2 B^2 B' B''' - (B B'')^2 - [B']^4 ) )\right)\\
f & = & \frac{1}{4}\int\frac{d^4q}{(2\pi)^4}\;K^2\left(
  (B B')^2+ q^2 B B' ( [B']^2 + B B'')
           + \frac{1}{6}\, q^4\,  ( [B']^2 + B B'')^2  \right)\\
g & = & \frac{1}{12}\,\int\frac{d^4q}{(2\pi)^4}\;
        q^4\,K^2\,\left( [B']^2 + B B''\right)^2 \\
h & = & \frac{1}{4}\,\int\frac{d^4q}{(2\pi)^4}\;
        q^2 K^2 B B' \left( [B']^2 + B B''
                + \frac{1}{3} q^2 ( 3 B' B'' + B B''')\right)\\
i & = & -\int\frac{d^4q}{(2\pi)^4}\;q^2\,K^2\,B^2\left(\rule{0mm}{5mm}
         A B'( A\,B\,K'\, + 2\,B\,A'\,K\,
        + A\,B'\,K) \right.+\frac{1}{3}\,q^2\,\left[\rule{0mm}{5mm} \right.
        {4\,A\,B\,A'\,B'\,K'\,} \nonumber \\
& &  \;\;\;\;\;\;\;\;\;\;\;\;\;\;\;\; + {8\,B\,[A']^2\,B'\,K} +
        {5\,A\,A'\,[B']^2\,K} +
        {2\,A\,B\,B'\,{\it A''}\,K} +
        {A\,B\,A'\,{\it B''}\,K} \nonumber \\
&  & \left.\left.\;\;\;\;\;\;\;\;\;\;\;\;\;\;\;\;
+q^2\,A'({4\,{B}\,A'\,B'\,K'\,} + {5\,A'\,[B']^2\,K\,} +
  {4\,{B}\,B'\,{\it A''}\,K} + {{B}\,A'\,{\it B''}\,K\,})\right]\right)\\
j & = & \int\frac{d^4q}{(2\pi)^4}\; K^2 B^2 \left(\rule{0mm}{5mm}A^2 B B' K
\right. \nonumber \\
& &  \;\;\;\;\;\;\;\;\;\;\;\;\;\;\;\;
 +q^2\left[ \frac{1}{2}A\,K\,(2 A' B B' + A \{ [B']^2 + B B''\})
        +\frac{1}{3} q^2\,\left\{\rule{0mm}{5mm} 4 A B A' B' K'
\right. \right. \nonumber\\
& & \;\;\;\;\;\;\;\;\;\;\;\;\;\;\;\;
+ 5 B [A']^2 B' K + 5 A A' [B']^2 K + 2 A B B' A'' K + A B A' B'' K
\nonumber\\
&& \;\;\;\;\;\;\;\;\;\;\;\;\;\;\;\;
\left.\left.\left.+ q^2 A'
        (4 B A' B' K' + 5 A' [B']^2 K + 4 B B' A'' K
                + B A' B'' K )\right\}\right]\right)\\
k & = & \frac{1}{4}\,\int\frac{d^4q}{(2\pi)^4}\; q^2 K^3 B^3 [B']^3 \\
l & = & \frac{1}{3}\,\int\frac{d^4q}{(2\pi)^4}\;
        \left( q^4 K^3 B^2 [B']^2 ( [B']^2 + B B'' )
                + \frac{1}{2}\, q^2 \,K^3 B^3 [B']^3 \right)\\
m & = & \frac{1}{3}\,\int\frac{d^4q}{(2\pi)^4}\;q^4 K^4 B^4 [B']^4 )\\
n_{K_2} & = & \int\frac{d^4q}{(2\pi)^4}\; q^2 K^4 B^4 [B']^2
                \left(A^2 + \frac{2}{3}\,q^2 (A A' + q^2[A']^2) \right) \\
n_{K_3} & = & \int\frac{d^4q}{(2\pi)^4}\; q^2 K^4 B^4 [B']^2
                \left(A^2 - \frac{2}{3}\,q^2 (A A' + q^2[A']^2) \right)
\end{eqnarray}
In these expressions
\begin{equation}
F' = \frac{d}{dq^2} F(q^2)
\end{equation}
with $F''$ representing the second derivative, etc.

The $\pi$ charge radius obtained from the generalised impulse approximation
diagram, Fig.~\ref{RpiGIA}, is
\begin{equation}
\label{Rpi}
\langle r_{\pi}^2\rangle =
\frac{N_c}{2 f_\pi^2}\int\frac{d^4q}{(2\pi)^4}\, H(q^2)
\end{equation}
where
\begin{eqnarray}
H(x=q^2) & = &
{\it B}\,\left (-{\it A}^{2}{\it B}^{2}{\it B'}\,{\it K''}\,{\it
K}^{2}x^{2}-{\it A}^{4}{\it B''}\,{\it K'}\,{\it K}^{2}x^{3} -36\,
{\it A}^{2}{\it B}^{3}{\it K'}\,{\it K}^{2}-{\it A}^{4}{\it B'}
\,{\it K''}\,{\it K}^{2}x^{3}\right. \nonumber \\
& & +{\it A}^{2}{\it B}\,{\it A'}\,{\it
A''}\,{\it K}^{3}x^{3}-30\,{\it A}\,{\it B}^{3}{\it A'}\,{\it K}
^{3} -42\,{\it A}^{2}{\it B}^{2}{\it B'}\,{\it K}^{3}-24\,{\it A}^
{2}{\it B}^{3}{\it K''}\,{\it K}^{2}x \nonumber \\
& & -48\,{\it A}^{4}{\it B}\,{
\it K'}\,{\it K}^{2}x-54\,{\it A}\,{\it B}^{3}{\it A'}\,{\it K'}
\,{\it K}^{2}x-42\,{\it A}^{2}{\it B}^{2}{\it B'}\,{\it K'}\,{
\it K}^{2}x -72\,{\it A}^{3}{\it B}\,{\it A'}\,{\it K}^{3}x \nonumber \\
& & -18\,{
\it B}^{3}{\it A'}^{2}{\it K}^{3}x-24\,{\it A}^{4}{\it B'}\,{\it
K}^{3}x-42\,{\it A}\,{\it B}^{2}{\it A'}\,{\it B'}\,{\it K}^{3}x
-24\,{\it A}\,{\it B}^{3}{\it A''}\,{\it K}^{3}x \nonumber \\
& & -21\,{\it A}^{2}{
\it B}^{2}{\it B''}\,{\it K}^{3}x-4\,{\it A}^{2}{\it B}^{3}{\it
K'}^{3}x^{2}+4\,{\it A}^{2}{\it B}^{3}{\it K''}\,{\it K'}\,{\it
K}\,x^{2}-6\,{\it A}\,{\it B}^{3}{\it A'}\,{\it K'}^{2}{\it K}\,
x^{2} \nonumber \\
& & -2\,{\it A}^{2}{\it B}^{2}{\it B'}\,{\it K'}^{2}{\it K}\,x^{
2}-24\,{\it A}^{4}{\it B}\,{\it K''}\,{\it K}^{2}x^{2}-3\,{\it A}
\,{\it B}^{3}{\it A'}\,{\it K''}\,{\it K}^{2}x^{2}-68\,{\it A}^{3
}{\it B}\,{\it A'}\,{\it K'}\,{\it K}^{2}x^{2} \nonumber \\
& & -2\,{\it B}^{3}{
\it A'}^{2}{\it K'}\,{\it K}^{2}x^{2}-28\,{\it A}^{4}{\it B'}\,{
\it K'}\,{\it K}^{2}x^{2}-18\,{\it A}\,{\it B}^{2}{\it A'}\,{\it
B'}\,{\it K'}\,{\it K}^{2}x^{2}-8\,{\it A}\,{\it B}^{3}{\it A''}
\,{\it K'}\,{\it K}^{2}x^{2} \nonumber \\
& & +3\,{\it A}^{2}{\it B}^{2}{\it B''}\,
{\it K'}\,{\it K}^{2}x^{2}-20\,{\it A}^{2}{\it B}\,{\it A'}^{2}{
\it K}^{3}x^{2}-40\,{\it A}^{3}{\it A'}\,{\it B'}\,{\it K}^{3}x^{
2}-14\,{\it A}\,{\it B}\,{\it A'}\,{\it B'}^{2}{\it K}^{3}x^{2}\nonumber \\
& & +2
\,{\it A}^{2}{\it B'}^{3}{\it K}^{3}x^{2}-36\,{\it A}^{3}{\it B}
\,{\it A''}\,{\it K}^{3}x^{2}-5\,{\it B}^{3}{\it A'}\,{\it A''}\,
{\it K}^{3}x^{2}-4\,{\it A}\,{\it B}^{2}{\it B'}\,{\it A''}\,{
\it K}^{3}x^{2} \nonumber \\
& & -8\,{\it A}^{4}{\it B''}\,{\it K}^{3}x^{2}+2\,{\it
A}\,{\it B}^{2}{\it A'}\,{\it B''}\,{\it K}^{3}x^{2}+4\,{\it A}^
{2}{\it B}\,{\it B'}\,{\it B''}\,{\it K}^{3}x^{2}-4\,{\it A}^{4}{
\it B}\,{\it K'}^{3}x^{3}\nonumber \\
& & +4\,{\it A}^{4}{\it B}\,{\it K''}\,{\it
K'}\,{\it K}\,x^{3}-6\,{\it A}^{3}{\it B}\,{\it A'}\,{\it K'}^{2
}{\it K}\,x^{3}-2\,{\it A}^{4}{\it B'}\,{\it K'}^{2}{\it K}\,x^{3
}-3\,{\it A}^{3}{\it B}\,{\it A'}\,{\it K''}\,{\it K}^{2}x^{3}\nonumber\\
& & -10
\,{\it A}^{2}{\it B}\,{\it A'}^{2}{\it K'}\,{\it K}^{2}x^{3}-10\,
{\it A}^{3}{\it A'}\,{\it B'}\,{\it K'}\,{\it K}^{2}x^{3}-4\,{
\it A}^{3}{\it B}\,{\it A''}\,{\it K'}\,{\it K}^{2}x^{3}-6\,{\it
A}\,{\it B}\,{\it A'}^{3}{\it K}^{3}x^{3} \nonumber \\
&& \left. -6\,{\it A}^{2}{\it A'}
^{2}{\it B'}\,{\it K}^{3}x^{3}-2\,{\it A}^{3}{\it B'}\,{\it A''}
\,{\it K}^{3}x^{3}-2\,{\it A}^{3}{\it A'}\,{\it B''}\,{\it K}^{3}
x^{3}. \right )
\label{ChRad}
\end{eqnarray}
which is obtained with the fermion-photon vertex Ansatz of Ref.~\cite{BPR92}:
\begin{eqnarray}
\lefteqn{\Gamma_\mu (p,q)  =
\frac{1}{2} [A(p) + A(q)]\gamma_\mu } \label{eq:vtx} \\
& & +\frac{(p+q)_\mu}{p^2 -q^2}
\left\{ [A(p)-A(q)]\frac{1}{2}(\gamma\cdot p + \gamma\cdot q )
       -i [B(p) - B(q)] \right\}~.\nonumber
\end{eqnarray}
which ensures local current conservation in this calculation.

Computer code for all of the quantities listed in this appendix in
either MATHEMATICA or FORTRAN format is available upon request.


\figure{\label{qloop}This quark loop diagram contributes to $f_\pi$.  The
filled
circles represent $\langle\pi|\overline{q}q\rangle$ Bethe-Salpeter amplitudes,
the thick external lines represent the $\pi$ field and the thin internal lines
represent dressed quark propagators.}
\figure{\label{RpiGIA} This diagram provides the piece of $r_\pi$,
Eq.~(\ref{Rpi}), that is regular in the chiral limit.  The thick, straight
external lines represent the incoming and outgoing $\pi$, the filled circles at
the $\pi$ legs represent the $\langle\pi|\overline{q}q\rangle$ Bethe-Salpeter
amplitudes, the wiggly line represents the photon, $\gamma$, the shaded circle
at the $\gamma$ leg represents the regular part of the dressed quark-photon
vertex (which satisfies the Ward Identity), Eq.~(\ref{eq:vtx}), and the thin
internal lines represent the dressed quark propagator in the model.}
\figure{\label{PlotB} The function $B(p^2)$ of Eq.~(\ref{Bmodel}) obtained with
the parameters in Eq.~(\ref{aFit}): solid line; compared with that obtained
with
Eq.~(\ref{BurdenP}): dashed line.}
\figure{\label{PWA} The partial wave amplitudes of Eqs.~(\ref{T00}-\ref{T22})
obtained with the best fit described in connection with Eq.~(\ref{bBfit}):
solid line.  In the plots of \mbox{$T_0^0(x)$}, \mbox{$T_1^1(x)$} and
\mbox{$T_0^2(x)$} we show, for comparison, the results that would be obtained
with the Weinberg result for \mbox{$A(s,t,u)$} in Ref.~\cite{W66} (short-dashed
line): this form gives \mbox{$T_2^0(x)\equiv 0 \equiv T_2^2(x)$}.  In these
plots we also show the extrapolation to threshold of Roy equation fits to
high-energy $\pi$-$\pi$ scattering data for these partial waves: long-dashed
line.  The data are extracted from Fig.~4 in Ref.~\cite{DRV88}.  We have
defined \mbox{$x= E^2 - 4 m_\pi^2$}.  }
\begin{table}
\caption{In this table we summarise our results for the scattering lengths
calculated in Secs.~\ref{scatfit} and~\ref{scatcalc}.  The column labelled
``Exp'' lists the ``model experimental data'' set that we constructed; see
Eq.~(\ref{ModelExp}).  Fit~1 is the best fit to ``Exp'' possible (in the sense
of minimising the sum of the squares of the differences) in the GCM when the
$K$s are allowed to vary; see Eq.~(\ref{bestas}).  Fit~2 is the best fit
possible when one sets $K_1 = 0$ and allows $K_{2,3}$ to vary and corresponds
to
the best fit to ``Exp'' possible with the parametrisation of Ref.~\cite{DRV88};
see Eq.~(\ref{bestasD}).  Cal.~1 is the calculated result obtained in the point
meson limit of the GCM and corresponds to the result that would be obtained in
the Nambu--Jona-Lasinio model [\mbox{$1/(96\pi^2)=0.00106$}]; see
Eq.~(\ref{PML}). Cal.~2 is the calculated result using the quark propagator
given by Eqs.~(\ref{Amodel}) and (\ref{Bmodel}) and using the parameters of
Eq.~(\ref{aFit}); see Eq.~(\ref{bBfit}).}
\pagebreak
\begin{tabular}{||c|l|l|l|l|l||}
$K_1=$ & & -0.000455  & 0.0  & $\frac{1}{96\pi^2}$ &   0.000460 \\
$K_2=$ & &  \ 0.000414  & 0.000844 &$\frac{1}{96\pi^2}$& 0.00104 \\
$K_3=$ & &  \ 0.00150  & 0.00149 & $\frac{1}{96\pi^2}$ & 0.000868 \\\hline
      & Exp &   Fit 1 & Fit 2 & Cal. 1 & Cal. 2\\\hline
$a_{0}^0$ &\  0.200   &\  0.15    &\  0.15 &\  0.17    &\  0.17\\\hline
$a_{0}^2$ & -0.037  & -0.042  & -0.045 & -0.048& -0.052\\\hline
$a_{1}^1$ &\   0.038 &\  0.035   &\  0.037 &\  0.036  &\  0.032\\\hline
$a_{2}^0$ &\   0.0017&\  0.0017  &\  0.0018 &\  0.0020&\  0.0011\\\hline
$a_{2}^2$ & -0.0011& -0.0011 & -0.0011 & \ 0.0   & -0.00085\\\hline
$ \frac{1}{5} \sum_{i=1}^{5}
        \left|\frac{a_{\rm Fit}^i}{a_{\rm ``Exp''}^i} -1\right|$ &
        & \ 0.10 & \ 0.11 & \ 0.34& \ 0.20\\
\end{tabular}
\end{table}
\end{small}
\end{document}